\documentstyle{article}
\begin{document}
\normalsize
\centerline{\bf THE LEONTOVICH BOUNDARY CONDITIONS AND}
\centerline{\bf CALCULATION OF EFFECTIVE IMPEDANCE OF INHOMOGENEOUS 
METAL} 
\normalsize
\vskip 5mm
\centerline{\it Alexandr M. Dykhne}
\small
\centerline{TRINITI}
\centerline{142190 Troitsk, Moscow Region; 
e-mail: dykhne@fly.triniti.troitsk.ru}
\vskip 3mm
\centerline{\it Inna M. Kaganova}
\small
\centerline{Institute for High Pressure Physics Russian Academy of 
Sciences}
\centerline{142190 Troitsk, Moscow Region; e-mail: 
litin@hppi.troitsk.ru}
\normalsize
\begin{abstract}
We bring forward rather simple algorithm allowing us to calculate the 
effective impedance of inhomogeneous metals in the frequency region   
where the local Leontovich (the impedance) boundary conditions are 
justified. The inhomogeneity is due to the properties of the metal 
or/and the surface roughness. Our results are nonperturbative ones with 
respect to the inhomogeneity amplitude. They are based on the recently 
obtained exact result for the effective impedance of inhomogeneous 
metals with flat surfaces \cite{1,2}. One-dimension surfaces 
inhomogeneities are examined. Particular attention is paid to the 
influence of generated evanescent waves on the reflection 
characteristics. We show that if the surface roughness is rather strong, 
the element of the effective impedance tensor relating to the p-
polarization state is much greater than the input local impedance. As 
examples, we calculate: i) the effective impedance for a flat surface 
with strongly nonhomogeneous periodic strip-like local impedance; ii) 
the effective impedance associated with one-dimensional lamellar 
grating. For the problem (i) we also present equations for the forth 
lines of the Pointing vector in the vicinity of the surface.
\end{abstract} 



\section {INTRODUCTION}

It is well known that to solve an electrodynamic problem external with 
respect to a metal, one can use the impedance boundary conditions 
\cite{3}
$$
{\bf E}_{t} = \hat \zeta [{\bf n},{\bf H}_{t}],    \eqno (1)
$$
${\bf E}_{t}$ and ${\bf H}_{t}$ are the tangential components of 
electric and magnetic vectors at the metallic surface; $\bf n$ is the 
unit external normal vector to the surface of the metal; $\hat\zeta$ is  
the two-dimensional surface impedance tensor. Its real and imaginary 
parts define the dissipated energy and the phase shift of the reflected 
electromagnetic wave respectively (see, for example, \cite {3}). 

If for an inhomogeneous metal the penetration depth of an 
electromagnetic wave $\delta$ is much less than a characteristic size 
$a$ of the surface inhomogeneity, omitting the terms of the order of 
$|zeta|\delta /a$, the local values of the impedance can be used in the 
boundary conditions (1). In this case Eqs.(1) are usually called the 
local Leontovich (the impedance) boundary conditions \cite{4}. The 
surface inhomogeneity can be due to the inhomogeneity of the properties 
of the metal or/and the surface roughness. The elements of the local 
surface impedance tensor of an inhomogeneous metal are functions of 
position on the surface. If the metal surface is rough, the components 
of the unit vector $\bf n$ depend on position too. The more complex is 
the geometry and composition of the surface, the more complex is the 
solution of the problem. When the surface inhomogeneity is strong, the 
solution requires cumbersome numerical calculations. 

The structure of electromagnetic fields in the immediate vicinity of the 
inhomogeneous metallic surface is very complex too. The fields above the 
surface are represented by incident, reflected, scattered and evanescent 
waves. However, sometimes it is sufficient to know the reflected 
electromagnetic wave only, not going into details of the field 
distribution near the surface.

In what follows, when calculating we assume that
$$
\delta \ll a \ll \lambda,  \eqno (2)
$$
where $\lambda = 2\pi c/\omega$ is the vacuum wavelength. When 
inequalities (2) are fulfilled, only the fields averaged over the 
surface inhomogeneities "survive" far (at the distance much greater than 
the vacuum wavelength) from a stochastically or periodically 
inhomogeneous metallic surface. To calculate these fields, it is 
convenient to introduce the effective surface impedance relating the 
tangential averaged fields at a plane close to the real metallic surface 
(or coinciding with this surface, if the surface of the metal is flat).
 
We define the effective impedance tensor with the aid of equations 
similar to Eqs.(1) for a homogeneous metal with the flat surface, let us 
say, $x_3=0$:
$$
<{\bf E}_{t}> = {\hat\zeta}_{ef}[{\bf e}_{3},<{\bf H}_{t}>]; \eqno (3)
$$
${\bf e}_{3}$ is the unit normal vector to this plane, $<:>$ denote an 
average over the plane $x_3=0$. In the case of a stochastically 
inhomogeneous surface the averaging is carried out over the ensemble of 
realizations of the inhomogeneities; if the inhomogeneity is a periodic 
one, the brackets denote the average over the period. If the tensor 
${\hat\zeta}_{ef}$ is known, when calculating the averaged 
electromagnetic fields we return to the reflection problem from a 
homogeneous flat surface.

The general way providing the correct result for the effective impedance 
is to derive equations for the averaged electromagnetic fields both in 
the metal and in the medium over the metal (we assume it to be vacuum), 
as well as the boundary conditions for these fields at the metal-vacuum 
interface. By proceeding from these equations the tensor 
${\hat\zeta}_{ef}$ can be calculated.   

The similar method is used rather often when calculating effective 
characteristics of inhomogeneous media. It goes back to the pioneering 
works of I.M.Lifshitz and his co-authors\footnote{The first one was 
\cite{5}}. However, mostly this regular method is applicable when the 
inhomogeneity amplitude is small and the perturbation theory can be used 
(with regard to the calculation of the effective impedance see 
\cite{6,7,8}). The exact solutions for effective characteristics of 
strongly inhomogeneous media can be found very rarely. One of examples 
is the calculation of the effective static conductivity of some two-
dimensional inhomogeneous metals \cite{9}.  

The other example is a quite new result for the effective surface 
impedance of an inhomogeneous metal with flat surface \cite{1,2}. It is 
valid in the frequency region of the local impedance boundary conditions 
applicability. Recently this solution was used when calculating the 
effective impedance of strongly anisotropic polycrystalline metals both 
under the conditions of normal \cite{1,2} and anomalous skin effect 
\cite{10}. 

In \cite{1,2} the expression for the effective impedance was obtained 
with the aid of the general scheme. Herein we present much more 
straightforward reasoning leading to the same result and extend it to 
the case of rough metallic surfaces. This result is a nonperturbative 
one with respect to the inhomogeneity amplitude. It is exact up to the 
limits of the local impedance boundary conditions (1) applicability. 

The organization of the paper is as follows. In Section 2 some general 
remarks useful when calculating the effective impedance are given. In 
Section 3 the simple way to derive the expression for ${\zeta}_{ef}$ in 
the case of flat inhomogeneous surfaces is presented. As an example, we 
analyze in details the case of one-dimensional strongly nonhomogeneous 
periodic strip-like local impedance. In Section 4 we present the 
algorithm allowing us to calculate the effective impedance in the case 
of one-dimensional rough surface. The obtained formulas are used to 
calculate the effective impedance associated with a lamellar grating. 
The s- and p-polarization states are examined separately. Concluding 
remarks are given in Section 5. 

\section {PRELIMINARY REMARKS}

Suppose the local impedance tensor $\hat\zeta$ is a known. A plane 
wave\footnote{To calculate the surface impedance of a metal it is 
sufficient to investigate the normal incidence of an electromagnetic 
wave onto the metal surface (see, for example, \cite{11})} ${\bf E}^{i} 
= {\bf E}_{0}{\rm e}^{-i(kx_3+\omega t)}$ is incident upon the surface 
from the region ${x}_{3}>0$. The metal occupies the volume beneath the 
plane ${x}_{3} = 0$. The surface of the metal is assumed to be the plane 
${x}_{3} = 0$ itself, or it is the plane ${x}_{3}=0$ perturbed by a 
roughness. For the sake of definiteness we suppose that the surfaces 
does not stand out over the plane $x_3=0$. Suppose the surface 
inhomogeneity is a stochastic or a periodic one. For a rough surface its 
profile is defined.    

Let $\bf E ({\bf x})$ and $\bf H ({\bf x})$ be the total electromagnetic 
vectors in the region $x_3 > 0$. We extract the averaged electric and 
magnetic vectors:
$$
\bf E({\bf x}) = <\bf E ({\bf x})> + {\bf e}(\bf x); \; 
\bf H({\bf x}) = <{\bf H}({\bf x})> + {\bf h}(\bf x); \;
<{\bf e}({\bf x})> = <{\bf h}({\bf x})>=0.
 \eqno (4.1);      
$$
If $a \ll \lambda$, the fields ${\bf e}({\bf x})$ and ${\bf h}({\bf x})$ 
represent mainly evanescent waves damping at the distances of the order 
of $a$ above the surface. When ${x}_3 \ge d$, where $d$ is a distance 
complying with the inequality $a\ll d \ll \lambda$, the leading terms in 
the expressions for the total fields are equal to their averaged values: 
$$
{\bf E} ({\bf x})\,\approx\,<{\bf E}(x_3)> = {\bf E}_{0}{e}^{-ik{x}_{3}} 
+ <{\bf E}^{r}> {e}^{ik{x}_{3}};\quad
{\bf H}\, \approx\, <{\bf H}(x_3)> = {\bf H}_{0}{e}^{-ik{x}_{3}} + <{\bf 
H}^{r}> {e}^{ik{x}_{3}};  \eqno (4.2)
$$ 
$<{\bf E}^{r}>$ and $<{\bf H}^{r}>$ are the electric and magnetic 
vectors of the reflected wave respectively. 

Suppose we are interested in the averaged fields only. Since no 
inhomogeneous parameters enter the Maxwell equations in the vacuum, the 
averaging of these equations provides the standard Maxwell equations for 
the averaged fields. Our problem is to obtain the boundary conditions 
for the averaged fields with the aid of the exact Eqs.(1). Comparing the 
last with Eqs.(3), we define the effective impedance. In what follows we 
solve this problem, reducing it to calculation of the magnetic field in 
the vicinity of infinitely conducting surface of the same geometry. 

To start the calculation, we remind that the elements of the tensor 
$\hat\zeta$ are of the order of the ratio $\delta/\lambda\ll 1$. 
Consequently, in accordance with Eqs.(1), the tangential components of 
electric vector at the metallic surface are much less then the 
tangential components of magnetic vector. 

Next, at the surface of a metal the magnetic vector ${\bf H}_{t}$ is 
approximately equal to the magnetic vector ${\bf H}^{per}_{t}$ at the 
surface of a perfect conductor ($|{\zeta}_{\alpha\beta}|=0$) of the same 
geometry. The same is true in the immediate vicinity of the metallic 
surface. Then at the plane ${x}_{3}=0$ 
$$
<{\bf H}_{t}>\, \approx \,<{\bf H}^{per}_{t}({\bf x}_{\parallel},0)>,
\eqno (5) 
$$
${\bf x}_{\parallel}$ is two-dimensional position vector at the plane 
${x}_{3}=0$. With regard to Eqs.(1) the first nonvanishing term in the 
expression for the tangential electric vector ${\bf E}_{t}$ at the 
metallic surface is
$$
{\bf E}_{t} \approx \hat\zeta ({\bf x})[{\bf n}({\bf x}),{\bf 
H}^{per}_{t}].  \eqno (6.1)
$$
If the surface is flat (it is the plane ${x}_{3}=0$ itself; ${\bf n} = 
{\bf e}_{3}$), the vector $<{\bf E}_{t}>$ is obtained from Eq.(6.1) 
directly: 
$$
<{\bf E}_{t}>\, = \,<\hat\zeta ({\bf x}_{\parallel})
[{\bf e}_3,{\bf H}^{per}_{t}]>;  \eqno (6.2)
$$
In \cite{2} we showed that the corrections to the expressions (6.2) are 
of the order of ${\delta}^{2}/a\lambda{\bf E}_0$. These small terms 
cannot be taken into account in the framework of the local Leontovich 
boundary conditions and, consequently, the averaged electric field 
cannot be calculated more accurately.

In the case of a flat inhomogeneous metallic surface Eqs.(3), (5) and 
(6.2) allow us to calculate the effective impedance. The obtained result 
cannot be improved. It is exact up to the limits of local impedance 
boundary conditions applicability.

The equations (5) and (6.1) allow us to define the effective impedance 
for a rough metallic surface too. However, the method of calculation is 
not so straightforward (see Section 4).

\section {FLAT SURFACE OF INHOMOGENEOUS METAL}

When the aforesaid argumentation is taken into account, the result of 
\cite{1,2} for the effective impedance of an inhomogeneous metal with a 
flat surface,
$$
{\hat\zeta}_{ef} = <\hat\zeta({\bf x}_{\parallel})>,   \eqno (7)
$$
is almost obvious. Indeed, in the case of normal incidence of an 
electromagnetic wave upon the flat surface ${x}_{3} = 0$ of a perfect 
conductor the magnetic vector at the surface is ${\bf H}_{t}^{per} = 
2{\bf H}_{0}$, where ${\bf H}_{0}$ is the amplitude of the magnetic 
field of the incident wave. This vector does not depend on position at 
the surface. Thus, with the aid of Eqs.(5) and (6.2) we have:  
$$
<{\bf H}_{t}> = 2{\bf H}_{0}; \quad
<{\bf E}_{t}> = 2<\hat\zeta>[{\bf e}_{3},{\bf H}_{0}].   \eqno (8)
$$
Comparing Eqs.(8) with Eqs.(3), we obtain Eq.(7). This result is correct 
for any metal with inhomogeneous flat surface both under the conditions 
of normal and anomalous skin effect. 

\subsection {ONE-DIMENSIONAL STRIP-LIKE LOCAL IMPEDANCE}

Suppose the local impedance tensor of an isotropic inhomogeneous metal 
is one-dimensional periodic strip-like function: 
${\hat\zeta}_{ik}({\bf x}_{\parallel}) = \zeta (x_1){\delta}_{ik}$ 
($i,k = 1,2 $), where 
$$
\begin{array}{rcl}
\zeta (x_1)& = & {\zeta}_{1}; \quad |x_1| < a,\\
&& {\zeta}_{2}; \quad a < |x_1| < b,
\end{array}
\eqno (9)
$$
$2b$ is the period. All the parameters, namely, $a$, $b$, $b-a$ and the 
local impedances is ${\zeta}_{1}$, ${\zeta}_{2}$ are compatible with 
inequalities (2).
 
We use this example not only to illustrate how simple the effective 
impedance can be calculated with the aid of Eq.(7), but also to 
visualize the difference between the fields in the immediate vicinity of 
one-dimensional inhomogeneous metallic surface for the s- and p-
polarization states. 

With regard to Eq.(7) for both polarizations the effective impedance is
$$
{\zeta}_{ef} = {\zeta}_{1}\frac{a}{b} + {\zeta}_{2}(1-\frac{a}{b}).
\eqno (10)
$$
However, the difference between the polarizations exhibits itself in the 
values of the fields near the surface. To complete the picture, for both 
polarizations we calculate the fields above the metal and the time-
averaged Pointing vector $\overline{{\bf S}} = c\overline{[{\bf E},{\bf 
H}]}/4\pi$ defining the time-averaged energy flux. The continuity 
equation for this vector, ${\rm div}\overline{{\bf S}}=0$, represents 
the energy conservation law (see, for example, \cite{3}). 

\underline{P-POLARIZATION}

Let the magnetic vector of the incident wave be ${\bf H}^{i} 
\left(0,{H}_{0},0\right){\rm exp}[-ik{x}_{3}]$. The only nonzero 
component of the total magnetic vector above the metal can be written as 
$$
H_2(x_1,x_3) = H_0\left\{{\rm e}^{-ik{x}_{3}}+ h_0{\rm e}^{ik{x}_{3}}+
h(x_1,x_3)\right\}, \eqno (11.1)
$$
where $h_0$ is the amplitude of the reflected wave and
$$
h(x_1,x_3) = \sum\limits_{-\infty ;q\ne 0 }^{\infty}h_q {\rm e}^{(i\pi 
qx_1/b - {\alpha}_{q}x_3)};\quad {\alpha}_{q} = \sqrt{{(q\pi/b)}^{2} - 
k^2} \approx \frac{q\pi}{b} . \eqno (11.2)
$$ 
represents the evanescent waves. With regard to the Maxwell equations 
the components of the electric vector are
$$
E_1(x_1,x_3) = H_0\left\{\frac{1}{k}\sum\limits_{-\infty ;q\ne 
0}^{\infty}
h_q{\alpha}_{q}{\rm e}^{(i\pi qx_1/b - {\alpha}_{q}x_3)} + h_0{\rm 
e}^{(ikx_3)} - {\rm e}^{(-ikx_3)}\right\};  \eqno (12.1)
$$
$$
E_3(x_1,x_3)=-H_0\frac{\pi}{kb}\sum\limits_{-\infty ;q\ne 0}^{\infty} 
qh_q{\rm e}^{(i\pi qx_1/b - {\alpha}_{q}x_3)}.   \eqno (12.2)
$$
We see that to calculate the components of the electric vector ${\bf 
E}(x_1,x_3)$ up to the terms of the order of $\zeta$ allowed in the 
framework of the local impedance boundary conditions applicability, we 
need to know the amplitudes of the evanescent waves $h_q$ ($q \ne 0)$ up 
to the terms of the order $(kb)\zeta \ll \zeta$.

According to Eqs.(1) in the p-polarization state $E_1(x_1,0) = - 
\zeta(x_1)H_2(x_1,0)$. We Fourier analyze this equation making use of 
Eqs.(11) and (12.1). Then, omitting the terms of the order of 
${\zeta}^{2}$, we obtain the coefficients $h_q$:  
$$
h_0 = 1 - 2{\zeta}_{ef};\quad 
h_q{|}_{q\ne 0}=\frac{2i}{\pi|q|}(ka)({\zeta}_{1} - {\zeta}_{2})F(\pi 
qa/b),
\eqno (13)
$$
where $F(x) = \sin x/x$. As a result up to the terms of the order of 
$\zeta$
$$
H_2(x,y) = 2H_0\left\{\cos (kby/\pi ) - {\zeta}_{ef}{\rm 
e}^{(ikby/\pi)}\right\};   \eqno (14.1)
$$
$$
E_1(x,y) = 2H_0\left\{i\sin (kby/\pi ) - 
{\zeta}_{ef}{\rm e}^{i(kby/\pi)} - 
\frac{2}{\pi}({\zeta}_{1}-{\zeta}_{2})C_1(x,y)\right\};   \eqno (14.2)
$$
$$
E_3(x,y) = \frac{4H_0}{\pi}({\zeta}_{1}-{\zeta}_{2})S_1(x,y).  \eqno 
(14.3)
$$
We introduce $x = \pi x_1/b$ and $y = \pi x_3/b$. The functions 
$C_m(x,y)$ and $S_m(x,y)$ are respectively the real and the imaginary 
parts of the function\footnote{It is easy to calculate the explicit form 
of the functions $R_m(z)$ for an arbitrary $m$, but trying to be short 
we do not present here the relevant expressions.} $R_m$ of the complex 
argument $z = x +iy$:
$$
R_m(z) = \sum\limits_{q=1}^{\infty}\frac{\sin \pi qa/b}{q^m}{e}^{iqz}. 
\eqno (15)
$$

Note, all the corrections taking into account the finite conductivity of 
the metal are of the order of the input impedances ${\zeta}_1$ and 
${\zeta}_{2}$. Next, the expression for the magnetic field, Eq.(14.1), 
is just the same as if calculated for the p-polarized wave incident upon 
a homogeneous metallic surface with the impedance ${\zeta}_{ef}$. 
However, the evanescent waves contribute to the electric field. In 
equations (14.2) and (14.3) they are represented by the terms including 
the functions $C_1(x,y)$ and $S_1(x,y)$. These terms are exponentially 
small when $y\gg 1$ (or $x_3 \gg b$).

From Eqs.(14) it follows that for the p-polarization state up to the 
terms of the order of $\zeta$ the Pointing vector $\overline{{\bf 
S}^{(p)}}$(the superscript $(p)$ indicates the polarization) can be 
written as
$$
\overline{{\bf S}^{(p)}}(x,y) = - \frac{c{|H_0|}^{2}}{\pi} 
\bigtriangledown F_p (x,y), \eqno (16)
$$
where
$$
F_p(x,y) = \frac{y}{2}{\rm Re}{\zeta}_{ef} - \frac{{\rm Re}({\zeta}_{1}-
{\zeta}_{2})}{\pi}\cos (kby/\pi )C_2(x,y). \eqno (17)
$$
Far from the surface ($y\gg 1$) the component $\overline{{S}_{1}^{(p)}}$ 
of the Pointing vector is exponentially small, and the component 
$\overline{{S}_{3}^{(p)}} = -(c/2\pi ){|H_0|}^{2}{\rm Re}{\zeta}_{ef}$. 
At the surface $\overline{{S}_{1}^{(p)}}=0$ and the component 
$\overline{{S}_{3}^{(p)}}$ is defined by the local impedance:
$\overline{{S}_{3}^{(p)}} = -(c/2\pi){|H_0|}^{2}{\rm Re}{\zeta}_{1}$, if
$|x_1| < a$ and $\overline{{S}_{3}^{(p)}} = -(c/2\pi){|H_0|}^{2}{\rm 
Re}{\zeta}_{2}$, if $a<|x_1|<b$.

It is easy to see that along with Eq.(16) the components of the vector 
${\overline {\bf S}^{(p)}}$ can be written as
$$
{\overline {\bf S}^{(p)}} = \frac{c{|H_0|}^{2}}{\pi}{\rm rot}{\bf 
A}_{p}(x,y), 
\eqno (18.1)
$$ 
where the vector ${\bf A}_{p} = -(0,A_p,0)$ and
$$
A_p(x,y) = \frac{x}{2}{\rm Re}{\zeta}_{ef} + \frac{{\rm Re}({\zeta}_{1} 
- {\zeta}_{2})}{\pi}\cos (kby/\pi )S_2(x,y). \eqno (18.2)
$$

From Eq.(16) it follows that in the p-polarization state in addition to 
equation ${\rm div}\overline {{\bf S}^{(p)}}=0$ we also have ${\rm 
rot}\overline {{\bf S}^{(p)}}=0$. Thus, in the p-polarization state the 
vector $\overline {{\bf S}^{(p)}}$ satisfies the same set of equations 
as the electric vector in two-dimensional electrostatic problems. Making 
use of this analogy we introduce a complex potential $w(z) = F_p(x,y) - 
iA_p(x,y)$ for the Pointing vector $\overline{{\bf S}^{(p)}}$:
$$
w(z) = -\frac{c{|H_0|}^{2}}{\pi}\left\{\frac{iz}{2}{\rm Re}{\zeta}_{ef} 
+ {\rm Re}({\zeta}_{1}-{\zeta}_{2})R_2(z)\right\}. \eqno (19)
$$
At the plane $(x,y)$ equations ${\rm Re}w(z)= {\rm const}$ define the 
equipotential lines for the vector $\overline{{\bf S}^{(p)}}$, and 
equations ${\rm Im}w(z)= {\rm const}$ define the force lines of this 
vector. Evidently, far from the surface ($y \gg 1$) the force lines are 
parallel to the $y$ axis. Near the surface ($y \le 1$) they are 
distorted due to the influence of evanescent waves. 

As an illustration, let us examine the limiting case of very strong 
inhomogeneity: $a/b \ll 1$, $|{\zeta}_{1}|/|{\zeta}_{2}|\gg 1$ and 
$(a/b)|{\zeta}_{1}|\gg |{\zeta}_{2}|$. With regard to Eq.(10) the 
relevant effective impedance is ${\zeta}_{ef} = (a/b){\zeta}_{1}$, and 
according to Eq.(19) the complex potential is
$$
w(z) = \frac{c{|H_0|}^{2}}{\pi}{\rm Re}{\zeta}_{ef}\left\{-\frac{i}{2}z 
+ \ln (1 - {e}^{iz})\right\}. \eqno (20.1)
$$
The force lines corresponding to this potential are given by equations
$$
\tan (x/2) = \tan ({x}_{\infty}/2)\frac{1-{e}^{-y}}{1+{e}^{-y}},
\eqno (20.2)
$$
where ${x}_{\infty}$ defines the position of the given line far from the 
surface. These force lines are shown in Fig.1 (full lines). All the 
lines being distributed uniformly far from the surface, meet at the 
point $x=0$ on the surface. Immediately near the surface the force lines 
are the straight lines $y = \alpha x$ with $\alpha = \cot 
({x}_{\infty}/2)$.

It can be easily verified that the complex potential defined by 
Eq.(20.1) is the  same as the electrostatic complex potential of the 
system of charged straight lines that are perpendicular to the plane 
$(x_1,x_3)$ and intersecting this plane at the points $x_1 = 2bn;\; 
n=0,\pm 1, \pm 2:...$.

\underline{S-POLARIZATION}

Let the electric vector of the incident wave be ${\bf E}^{i} = 
\left(0,{E}_{0},0\right){\rm exp}[-ik{x}_{3}]$. The total electric 
vector above the metal is
$$
E_2(x_1,x_3) = E_0\left\{{\rm e}^{-ik{x}_{3}}+ e_0{\rm e}{ik{x}_{3}}+
e(x_1,x_3)\right\},\qquad 
e(x_1,x_3) = \sum\limits_{-\infty; q\ne 0}^{\infty}
e_q {\rm e}^{(i\pi qx_1/b - {\alpha}_{q}x_3)}, \eqno (21)
$$
${\alpha}_{q}$ is defined in Eq.(11.2); $e_0$ corresponds to the 
reflected wave and $e(x_1,x_2)$ represents the evanescent waves.

To define the Fourier coefficients $e_0$ and $e_q$ we use the boundary 
conditions (1). Omitting the terms of the order of ${\zeta}^{2}/kb$, we 
obtain
$$
e_0 = -(1 - 2{\zeta}_{ef});\quad 
e_q{|}_{q\ne 0}=\frac{2a}{b}({\zeta}_{1} - {\zeta}_{2})F(\pi qa/b).
\eqno (22)
$$
(compare with Eq.(13)). Then in terms of the variables $x$ and $y$ our 
result for the fields above the surface is
$$
E_2(x,y) = 2E_0\left\{-i\sin (kby/\pi ) + {\zeta}_{ef}{e}^{ikby/\pi} + 
\frac{2}{\pi}({\zeta}_{1}-{\zeta}_{2})C_1(x,y)\right\},   \eqno (23.1)
$$
$$
H_1(x,y) = 2E_0\left\{\cos (kby/\pi) -{\zeta}_{ef}{e}^{ikby/\pi}-
\frac{2i({\zeta}_{1}-{\zeta}_{2})}{kb}C_0(x,y)\right\},   \eqno (23.2)
$$
$$
H_3(x,y) = 4iE_0\frac{({\zeta}_{1}-{\zeta}_{2})}{kb}S_0(x,y);   \eqno 
(23.3)
$$

In contrast to the p-polarization state the evanescent waves contribute 
both to the electric and magnetic fields near the metallic surface. 
Next, the contributions of evanescent waves to the components of the 
magnetic vector are of the order of $\zeta /kb \gg \zeta$ (compare with 
Eqs.(14)).

Of course, in accordance with the energy conservation law in the s-
polarization state the components of the time-averaged Pointing vector 
satisfy the equation ${\rm div} \overline{{\bf S}^{(s)}} = 0$. However, 
the curl of this vector is not equal to zero and, consequently, in the 
s-polarization state no complex potential can be introduced. 

To define the equation for the force lines of the vector $\overline{{\bf 
S}^{(s)}}$ we write it as
$$
\overline{{\bf S}^{(s)}} = \frac{c{|E_0|}^{2}}{\pi}{\rm rot}{\bf 
A}_s(x,y), \quad {\bf A}_{s} = -(0,A_s,0);\eqno (24)
$$ 
$$
A_s(x,y) = \frac{x}{2}{\rm Re}{\zeta}_{ef} + \frac{{\rm Re}({\zeta}_{1}-
{\zeta}_{2})}{kb}\left[\sin (kby/\pi)S_1(x,y)+ \frac{kb}{\pi}\cos 
(kby/\pi)S_2(x,y)\right] \eqno (25)
$$
(compare with Eqs.(18)). Then the force lines of the vector $\overline 
{{\bf S}^{(s)}}$ are given by equations $A_s = {\rm const}$. In the case 
of very strong inhomogeneity ($a/b \ll 1$, $|{\zeta}_{1}| \gg 
|{\zeta}_{2}|$ and $(a/b)|{\zeta}_{1}|\gg |{\zeta}_{2}|$), they are 
described by the equation
$$
\arctan u - 2\frac{u}{(1+u^2)}\frac{y{e}^{-y}}{(1-{e}^{-2y})} = 
\frac{\pi-{x}_{\infty}}{2},\qquad 
u(x,y) = \frac{1-{e}^{-y}}{1+{e}^{-y}}\cot (x/2)   \eqno (26.1)
$$
(compare with Eq.(20.2)). Again ${x}_{\infty}$ defines the position of 
the line far from the surface. Immediately near the surface the force 
lines are the straight lines $y={u}_{0}x$. For the given value 
${x}_{\infty}$ the slope $u_0$ is defined by the equation
$$
\arctan u_0 - \frac{u_0}{1+u_0^2}= \frac{\pi - {x}_{\infty}}{2}.  \eqno 
(26.2)
$$
These force lines are shown in Fig.1 (dashed lines).

Summarizing the results of this section we would like to emphasize that 
the main difference between the two polarizations is the presence of the 
terms of the order of $\zeta /kb \sim \delta /b $ in the expressions for 
the components of the magnetic vector in the s-polarization state. 

\section{1-D ROUGH SURFACE}

In this Section we show that if the metallic surface is rough, to 
calculate the effective impedance it is sufficient to know the magnetic 
vector ${\bf H}^{per}$ near the surface of a perfect conductor of the 
same geometry. The main problem is to calculate the components of this 
vector. In the general case they depend on the surface geometry and, as 
a rule, on position at the surface. 
 
We restrict ourselves with one-dimensional rough homogeneous metallic 
surfaces only. To be definite, let us suppose that our one-dimensional 
surface is a periodic one, $2b$ is the period. A part of the surface is 
the plane $x_3=0$ itself, and each period has a deepening with a throat 
of the length $2a$ (Fig.2a).

In the case of the p-polarization state we calculate the relevant 
element of the effective impedance tensor for an arbitrary surface 
profile. For the s-polarization state we perform the calculation for 
one-dimensional lamellar gratings. We use this example to show the 
difference between the polarizations. 

To show the method of calculation, let us suppose that we know the 
vector ${\bf H}^{per}$ for a given surface profile. With regard to the 
definition (3) of ${\zeta}_{ef}$ and Eq.(4.2) we have that far from the 
surface ($x_3 \gg \lambda$) the fields are the same as the ones above a 
flat metallic surface with the impedance equal to ${\zeta}_{ef}$. This 
means that far from the surface the only nonzero component of the time-
averaged Pointing vector is
$$
{\overline S}_{3} = -\frac{c{|H_0|}^{2}}{2\pi}{\rm Re}{\zeta}_{ef}
\quad (x_3 \gg \lambda), \eqno (27.1)
$$
where $H_0$ is the amplitude of the magnetic field in the incident wave. 
On the other hand, since near a metallic surface the magnetic vector 
${\bf H}$ is nearly the same as ${\bf H}^{per}$, at the rough metallic 
surface itself the time-averaged Pointing vector is
$$
{\overline {\bf S}} = - \frac{c{\rm Re}\zeta}{8\pi}
{|{\bf H}_{t}^{per}({\bf x})|}^{2}{\bf n}({\bf x}),  \eqno (27.2)
$$
$\bf n$ is the unit external normal vector to the surface (see, for 
example, \cite{3}). The energy conservation law requires ${\rm div} 
{\overline {\bf S}} = 0$, and, consequently, the energy fluxes through 
the infinitely distant surface and the surface of the metal are equal. 
Then
$$
{\rm Re}{\zeta}_{ef} = \frac{{\rm Re}\zeta}{8b{|H_0|}^{2}}
\int {|{\bf H}_{t}^{per}({\bf x})|}^{2}{\rm d}l,  
\eqno (27.3)
$$
the integration is carried out along one period of the line bounding the 
surface at the $(x_1,x_3)$ plane.

The last equation enables us to calculate the real part of the effective 
impedance. To calculate ${\rm Im}{\zeta}_{ef}$ we take into account that 
though no absorption happens when an electromagnetic wave reflects from 
a rough surface of the perfect conductor, a phase shift takes place. In 
other words, some pure imaginary surface impedance can be associated 
with the rough surface of the perfect conductor. It is easy to 
understand that its leading term is of the order of $kb$ (we remind that 
$kb \ll 1$). Since $|\zeta| \sim k\delta \ll kb$, in the case of the 
finite conductivity the same pure imaginary surface impedance defines 
the leading term in the expression for ${\rm Im}{\zeta}_{ef}$.

Now, supposing that $\zeta = 0$, with the aid of Eq.(3) we can define 
${\rm Im}{\zeta}_{ef}$ as the ratio of the tangential electric and 
magnetic fields averaged across the plane $x_3=0$. The result for the 
averaged tangential magnetic vector is rather obvious. Indeed, Eq.(3) 
have the form of the boundary conditions for a flat metallic surface. 
Consequently, the leading term in the expression for the averaged 
tangential magnetic field is $<{\bf H}_{t}> = 2{\bf H}_{0}$. This 
statement is true both for a metallic surface and the surface of the 
perfect conductor. Next, when $\zeta = 0$, only the throat of the 
deepening contributes to the averaged electric field: 
$$
<{\bf E}_{t}^{per}> = \frac{1}{2b}\int\limits_{-a}^{a} 
{\rm d}{x}_{1} \, {\bf E}_{t}^{per}({x}_{1},{x}_{3} = 0). \eqno (28) 
$$
If the magnetic vector ${\bf H}^{per}$ is known, the last integral can 
be calculated with the aid of the Stokes theorem or one of its 
modifications (see below). 

Concluding this subsection we would like to note, that the components of 
the vector ${\bf H}^{per}$ depend on the parameter $kb \ll 1$ (see 
Eq.(2)). We show that when calculating the leading terms of ${\rm 
Re}{\zeta}_{ef}$ and ${\rm Im}{\zeta}_{ef}$, we need to know the field 
${\bf H}^{per}$ up to the terms independent of this small parameter 
only. 

In what follows p- and s- polarization states are examined separately.

\subsection {CALCULATION OF ${\zeta}_{11}^{(ef)}$ (P-POLARIZATION 
STATE)}

Let us start with the p-polarization state (the magnetic vector ${\bf 
H}=(0,H_2,0)$ is parallel to the rulings). With regard to our notations 
the relevant element of the effective impedance tensor is 
${\zeta}_{11}^{ef}$. 

It is easy to see that when a p-polarized wave is incident upon one-
dimensional rough surface of a perfect conductor, and the characteristic 
sizes of the surface profile are compatible with inequalities (2), in 
the vicinity of the surface the expression for the magnetic field 
$H_2^{per}$ can be written as $H_2^{per} = 2H_0 + (kb)h(x_1,x_3;kb);\, 
kb \ll 1 $. (In APPENDIX 1 we calculate $H_2^{per}$ for an infinitely 
conducting lamellar grating.) As far as the terms of the order of 
$(kb)\zeta$ falls outside the framework of the local impedance boundary 
conditions applicability, the corrections $(kb)h(x_1,x_3;kb)$ has to be 
omitted when calculating ${\rm Re}{\zeta}_{ef}$ with the aid of 
Eq.(27.3).

Next, to calculate the averaged tangential electric field $<{\bf 
E}_{t}^{per}> = (<{E}_1^{(per)}>,0,0)$ entering Eq.(28), we write the 
Stokes theorem: 
$$
\oint{\rm d}{\bf S}[\nabla ,{\bf E}^{per}] = 
\oint{\rm d}{\bf r}{\bf E}^{per}. \eqno (29)
$$
The integral in the left-hand side of Eq.(29) is taken over the cross-
section of the deepening (${\rm d}{\bf S} = (0,{\rm d}S,0)$); the 
integral in the right-hand side is carried out in the clockwise 
direction over the contour and the throat of the deepening. Now, taking 
into account that on the surface of the perfect conductor the tangential 
electric field is equal to zero, with the aid of the Maxwell equations 
we obtain
$$
<{E}_1^{per}> = \frac{ik}{2b} \int {\rm d}S\, {H}_{2}^{per}(x_1,x_3).   
\eqno (30)
$$
The integral in the right-hand side of Eq.(30) is of the order of 
$kSH_2/b$, $S$ is the area of the deepening. Since according to Eqs.(2) 
the value of $kS/b \ll 1$, when calculating the leading term in the 
expression for $<({E}_1^{(per)}>$ we again use the approximation 
${H}_{2}^{per}(x_1,x_3) = 2H_0$.

Let ${L}_{c}$ be the length of the surface contour with respect to one 
period. Now, for an arbitrary shape of the deepening we can write the 
expression for the effective impedance leaving only the leading terms of 
${\rm Re}{\zeta}_{ef}$ and ${\rm Im}{\zeta}_{ef}$:
$$
{\zeta}_{11}^{(ef)} = {\rm Re}\zeta \frac{L_c}{2b} + i\frac{kS}{2b}.
\eqno (31) 
$$

We see that independently of the shape and the size of the deepening 
${\rm Re}{\zeta}_{11}^{ef} > {\rm Re}\zeta$ and ${\rm 
Im}{\zeta}_{11}^{ef}$, being of the order of $k$ multiplied by a 
characteristic size of the surface profile, is much greater than ${\rm 
Im}\zeta \sim k\delta$. From Eq.(31) it follows that in the p-
polarization state an intensification of absorption due to the surface 
roughness, is merely geometrical effect relating to an increase of the 
area of the absorbing surface.

Let us estimate the maximum value ${\zeta}_{max}$ of ${\rm Re} 
{\zeta}_{11}^{ef}$. It is evident that ${\zeta}_{max}$ corresponds to 
the maximum permissible value of the ratio $L_c/2b$. Taking account of 
the inequalities (2), suppose the length $L_c$ is of the order of the 
vacuum wavelength $\lambda$ and the period $2b$ is of the order of the 
penetration depth $\delta$. Then, since $|\zeta |\sim \delta/\lambda$, 
we have ${\zeta}_{max} \sim 1$. (We would like to recall that for good 
metals $|\zeta| \ll 1$.) Of course, this is only the upper bound of the 
permissible value of ${\rm Re}{\zeta}_{11}^{ef}$. However, the estimate 
shows that when the length of the contour is much greater then the 
period ($L_c \gg 2b$), the value of ${\rm Re}{\zeta}_{11}^{ef}$ exceeds 
${\rm Re}\zeta$ significantly. Such increase of the effective impedance 
has to manifest itself as giant absorption of the incident p-polarized 
wave.  

The surface with $L_c \gg 2b$ can be realized, for example, in such a 
way. Suppose a planar metallic surface has periodic grooves. Suppose the 
boundary of the groove is a branching line (Fig.2.b). Let us say, this 
line has a fractal structure. Then the ratio ${L}_{c}/2b$ can be very 
large and, consequently, the value of the ${\rm Re}{\zeta}_{11}^{(ef)}$ 
close to ${\zeta}_{max}$ can be achieved. When the throats of the 
grooves are sufficiently narrow, almost all absorption takes place 
inside "the pockets" beneath the plane $x_3=0$.

\subsection {CALCULATION OF ${\zeta}_{22}^{(ef)}$ (S-POLARIZED WAVES)}

It is much more difficult to calculate the effective impedance relating 
to the s-polarization state (the electric vector ${\bf E}=(0,E_2,0)$ is 
parallel to the rulings). In our notations this is the element 
${\zeta}_{22}^{ef}$ of the effective impedance tensor. The point is that 
for this polarization the magnetic vector ${\bf H}^{per}$ varies 
significantly in the vicinity of the surface. Inside the grooves the 
strength of the magnetic field exponentially decays when the distance 
from the plane ${x}_{3}=0$ increases. Next, in contrast to the p-
polarization state, evanescent waves generated in the region $x_3 > 0$, 
contribute to the leading term of the tangential magnetic field 
${H}_{1}^{per}$. As a result even for rather simple surfaces, such as 
lamellar grating, the value of ${\zeta}_{22}^{(ef)}$ can be calculated 
only numerically.

Let us write down the general expression for ${\zeta}_{22}^{(ef)}$. 
Suppose the surface of the metal is of the type shown in Fig.2a, and we 
know the vector ${\bf H}_{t}^{per}$ in the vicinity of the surface. Then 
we can calculate ${\rm Re}{\zeta}_{ef}$ with the aid of Eq.(27.3).

To calculate ${\rm Im}{\zeta}_{ef}$ we need to know the averaged 
tangential magnetic $<H_1^{per}(x_1,0)>$ and electric 
$<E_2^{per}(x_1,0)>$ fields. Although the components of the magnetic 
vector depend on coordinates, with regard to the aforementioned 
arguments we can use the approximation $<H_1^{per}(x_1,0)> = 2H_0$ 
($H_0$ is the amplitude of the magnetic vector of the incident wave). To 
calculate $<E_2^{per}(x_1,0)>$ we use the vector modification of the 
Stokes theorem \cite{14} for the vector ${\bf E}^{per}$:
$$
\oint[[{\rm d}{\bf S},\nabla],{\bf E}^{per}] = \oint [{\rm d}{\bf 
r},{\bf E}^{per}]. \eqno (32.1)
$$
The domains of integration are the same as when calculating the integral 
(29). Multiplying the vector equation (32.1) by the unit vector ${\bf 
e}_{3}$ with regard to the Maxwell equations and the boundary conditions 
for the vector ${\bf E}^{per}$ we obtain
$$
<E_2^{per}(x_1,0)> = -\frac{ik}{2b} \int {\rm d}S H_1^{per}(x_1,x_3),  
\eqno (32.2)
$$

Now it is clear that in the s-polarization state the the effective 
impedance is
$$
{\zeta}_{22}^{(ef)} = {\rm Re}\zeta \frac{L_c}{2b}Z - \frac{ikS}{2b}W, 
\eqno (33.1)
$$
where the leading terms of $Z$ and $W$ are
$$
Z = \frac{1}{4L_c{|H_0|}^{2}}
\int {|{\bf H}_{t}^{per}({\bf x})|}^{2}{\rm d }l, \quad
W = \frac{1}{2SH_0}\int {\rm d}S H_1(x_1,x_3).  \eqno (33.2)
$$
(compare with Eq.(31)). Again $L_c$ is the length of the contour 
bounding our surface with respect to one period, and $S$ is the area of 
the deepening. The factors $Z$ and $W$ show the difference between the 
polarizations.

\subsection{${\zeta}_{22}^{(ef)}$ ASSOSIATED WITH LAMELLAR GRATING}

There are some regular methods allowing us to calculate the magnetic 
vector at the surface of a perfect conductor in the small roughness 
limit. However, if the surface roughness is not small, analytical 
solutions can be obtained for some specific surfaces only. Therefore, as 
an example, we examine an infinitely conducting lamellar grating 
(Fig.2c). We assume that all the sizes of the contour, namely the period 
$2b$, the width of the rectangular grooves $2a$, the depth of the 
grooves $h$ as well as $b-a$, are in agreement with inequalities (2). 

The electromagnetic fields in the vicinity of infinitely conducting 
lamellar gratings have been examined by a lot of authors (see, for 
example, \cite{13}). Therefore we do not go into details of calculation, 
but present only the basic formulae used to obtain the result and some 
remarks relating to the computational procedure. 

To calculate the vector ${H}_{t}^{per}$, following \cite{13}, we present 
the electric field in the central groove ${E}_{2}^{per} = 
{E}_{2}^{(-)}({x}_{1},{x}_{3})$ ($|x_1|< a;\, -h < x_3 < 0$) as series 
of modal functions ${\phi}_{n}^{(s)}(x_1,x_3)$ that are the solutions of 
the Maxwell equations:
$$
{E}_{2}^{(-)}({x}_{1},{x}_{3}) = \sum_{n=0}^{\infty}{B}_{n}^{(s)}
{\phi}_{n}^{(s)}(x_1,x_3), \quad {\phi}_{n}^{(s)}(x_1,x_3) = 
\sin \frac{\pi n}{2a}({x}_{1} - a)\sin {\beta}_{n}(h + {x}_{3}),
\eqno (34.1)
$$
${\beta}^{2}_n = k^2 - {(\pi n/2a)}^{2}$ and ${\rm Re}{\beta}_{n},\,
{\rm Im}{\beta}_{n} > 0$. The representation incorporates the boundary 
conditions for perfect conductors on the vertical and bottom horizontal 
facets of the groove.
 
In the region ${x}_{3}>0$ we seek the field ${E}_{2}^{per} = 
{E}_{2}^{(+)}({x}_{1},{x}_{3})$ as 
$$
{E}_{2}^{(+)}({x}_{1},{x}_{3}) = \left\{{\rm e}^{-ikx_3} +
 \sum_{q = -\infty}^{\infty} {e}_{q}^{+}{\rm e}^{i(\pi q{x}_{1}/b + 
{\alpha}_{q}{x}_{3})}\right\};
\eqno (34.2)
$$ 
${\alpha}_{0} = k$ and ${\alpha}_{q} = i\sqrt{{(\pi q/b)}^{2}-k^2}$, if 
$q \ne 0$. The amplitude of the incident wave is equal to one, $e_0$ is 
the amplitude of the reflected wave and $e_q\,(q\ne 0)$ are assigned to 
evanescent waves.

Taking account of the fields periodicity, matching the electric vector 
at the plane ${x}_{3} = 0$ and the tangential magnetic vector at the 
throat of the central groove, after eliminating the amplitudes $e_0$ and 
$e_q$, we obtain a matrix equation for the coefficients ${B}_{2n-
1}^{(s)}$. It has been repeatedly shown (see, for example, \cite{13}) 
that in the case of the s-polarization state all the even coefficients 
$B_{2n}$ are equal to zero. 

Let us introduce a dimensionless parameter $\gamma = kb \ll 1$. From 
Eqs.(33) it follows that when calculating ${\zeta}_{22}^{(ef)}$ we need 
to know the components of the vector ${\bf H}^{per}$ up to the terms 
independent of $\gamma$ only. We seek the coefficients ${B}_{2n-
1}^{(s)}$ $(n = 1,2...)$ as series expansions in powers of $\gamma$. The 
first non-vanishing terms of these series are proportional to $\gamma$. 
To calculate the magnetic vector up to the terms independent of 
$\gamma$, we cut off the series at the first terms. In this 
approximation in place of ${B}_{2n-1}^{(s)}$ we introduce the 
coefficients $Y_n$:
$$
B_{2n-1}^{(s)} = \frac{\gamma Y_n}{(2n-1)\sinh {\chi}_{n}}; \quad 
{\chi}_{n} = \frac{\pi (2n-1)h}{2a}.  \eqno (35)
$$

The coefficients $Y_n$ are the solution of the infinite set of 
equations: 
$$
\sum\limits_{n=1}^{\infty}Y_n {\Delta}_ {nm}(\mu) + 
\frac{{\pi}^{2}\coth {\chi}_{n}}{4\mu (2m-1)}Y_m = \frac{2}{{(2m-
1)}^{2}};
\eqno (36.1)
$$
$$
{\Delta}_ {nm}(\mu) = \sum\limits_{q=1}^{\infty}
\frac{(q\mu )[1 + \cos q\pi\mu ]}{[{(q\mu)}^{2}-{(2n-
1)}^{2}][{(q\mu)}^{2}-{(2m-1)}^{2}]}; \quad \mu = 2a/b. \eqno (36.2)
$$

Next, with ${H}_{1}^{(+)}({x}_{1},{x}_{3})$ to denote 
${H}_{1}^{per}({x}_{1},{x}_{3})$ in the region above the plane 
${x}_{3}=0$, and ${\bf H}^{(-)}({x}_{1},{x}_{3})$ to denote ${\bf 
H}^{per}({x}_{1},{x}_{3})$ in the central groove, in terms of the 
coefficients $Y_n$ we have in the limit $\gamma \to 0$:
$$
{H}_{1}^{(+)}({x}_{1},0) = 2\left[1+ \delta H(x_1)\right],  \eqno (37.1)
$$
$$
\delta H(x_1) = \frac{2a}{b}\sum\limits_{q=1}^{\infty}
q\cos \left(\frac{\pi qx_1}{b}\right)\Phi \left(\frac{2a}{b}q\right), 
\quad \Phi (z) = \cos \left(\frac{\pi 
z}{2}\right)\sum\limits_{n=1}^{\infty}
\frac{Y_n}{[{z}^{2}-{(2n-1)}^{2}]}.   \eqno (37.2)
$$
Next,
$$
{H}_{1}^{(-)}({x}_{1},{x}_{3}) = -\frac{\pi b}{2a 
}\sum\limits_{n=1}^{\infty}
\frac{Y_n}{\sinh {\chi}_{n}}\sin \left(\frac{\pi(2n-1)}{2a}
(x_1-a)\right)\cosh {\chi}_{n}\left(1+\frac{x_3}{h})\right),   \eqno 
(37.3)
$$
$$
{H}_{3}^{(-)}({x}_{1},{x}_{3}) = \frac{\pi 
b}{2a}\sum\limits_{n=1}^{\infty} \frac{Y_n}{\sinh {\chi}_{n}}\cos 
\left(\frac{\pi(2n-1)}{2a}(x_1-a)\right) \sinh 
{\chi}_{n}\left(1+\frac{x_3}{h})\right).   \eqno (37.4)
$$
Note, from Eqs.(37.1) and (37.2) it immediately follows that $<\delta 
H(x_1)> = 0$ and $<{H}_{1}^{+}({x}_{1},0)>=2$.

Thus, at first we need to calculate the coefficient $Y_n$ solving 
numerically the infinite set of equations (36). Then we use Eqs.(37) to 
calculate the components of the magnetic vector and define the real and 
imaginary parts of the effective impedance with the aid of Eqs.(33).
We describe the computational algorithm in APPENDIX 2. 

Let us write the coefficient $Z$ entering Eq.(33.1) as
$$
Z = Z_{pl} + Z_s,   \eqno (38.1)   
$$
where $Z_{pl}$ and $Z_s$ are related respectively to the parts of the 
plane $x_3=0$ incorporated in the contour and the inner facets of the 
grooves. With regard to Eqs.(37) we have
$$
Z_{pl} = \frac{1}{(b+h)}\int\limits_{a}^{b}{|1 + \delta H(x_1)|}^{2}{\rm 
d}x_1;
\eqno (38.2)
$$
$$
Z_s = \frac{1}{4(b+h)}\left\{\int\limits_{-h}^{0}{|H_3^{(-
)}(a,x_3)|}^{2}
{\rm d}x_3 + \frac{1}{2}\int\limits_{0}^{a}{|H_1^{(-)}(x_1,-h)|}^{2}{\rm 
d}x_1
\right\}. \eqno (38.3)
$$
Since the explicit forms of $Z_{pl}$ and $Z_s$ in terms of the 
coefficients $Y_n$ are very lengthy, they are not presented here. The 
coefficient $W$ defines the imaginary part of the effective impedance. 
For our lamellar grating
$$
W = \frac{b}{\pi h}\sum\limits_{n=1}^{\infty}\frac{Y_n}{{(2n-1)}^{2}}.
\eqno (38.4)
$$

In Figs.3 and Fig.4 we display our numerical results for two different 
values of the grooves depth. We choose $h/2a = 0.1$ and $h/2a = 5$ to 
represent shallow and deep grooves respectively. The dots
correspond to $h/2a = 0.1$ and the stars to $h/2a = 5$.

In Fig.3a we plot the ratio ${\rm Re}{\zeta}_{22}^{(ef)}/ {\rm 
Re}{\zeta}$ versus $a/b$ showing that the presence of the grooves leads 
to an increase of absorption. Our calculations show that if $h/2a<1$, 
the value of $M(a/b)$ increases with an increase of the ratio $h/a$. 
However, the results for $M=M(a/b)$ are practically the same when 
$h/2a>1$. This means that an incident wave "understands" the grooves, 
whose depth is more than $2a$, as infinitely deep. In other words, for 
our lamellar grating the stars in Fig.3a define the maximal values 
of the real part of the effective impedance in the s-polarization state.

In Fig.3b for the same values of $h/2a$ we present the values of 
$Z={\rm Re}{\zeta}_{22}^{(ef)}/{\rm Re}{\zeta}_{11}^{(ef)}$ for different 
ratios $a/b$. The coefficient $Z$ defines the distinction between 
the values of the effective impedance for the s- and p-polarization 
states. In Fig.3c we show the ratios $Z_{pl}/Z_s$ versus $a/b$. This 
ratio defines the distribution of the absorbed energy between the 
horizontals sections $x_3=0$ and the inner surfaces of the grooves. In 
Fig.4 the coefficients $W={\rm Im}{\zeta}_{22}^{(ef)}/ {\rm 
Im}{\zeta}_{11}^{(ef)}$ versus $a/b$ are plotted.

Our results show that the same as for p-polarized waves, the presence of 
the grating leads to an increase of absorption of incident s-polarized 
waves. However, since $Z<1$, always the absorption of the p-polarized 
waves is more intensive. The more deep are the grooves, the more the 
difference between the polarizations. We also see (Fig.4) that the same 
as the factor $Z$, the factor $W$ entering the expression for the 
imaginary part of the effective impedance (see Eq.(33.1)) is less than 
one. However, in contrast to $Z$, it depends on the ratio $a/b$ only 
slightly.

We must note that a similar problem was solved in \cite{17} by 
L.A.Vainshtein, S.M.Zhurav, A.I.Sukov. The authors of \cite{17} examined 
a one-dimensional grating composed of semi-infinite plates (infinitely 
deep grooves) exposed to s-polarized normally incident electromagnetic 
wave. They also made use of the impedance boundary conditions (1) and 
with the aid of some other method based on the solution of the Wiener-
Hopf equation, calculated the reflection coefficient $R$. Then supposing 
that $|\zeta| /kb \ll 1$, in the framework of the perturbation theory 
the value of $R$ was defined up to the terms of the order of $\zeta$. 
The real part of the effective impedance was determined when calculating 
the difference $1-{|R|}^{2}$.

Of course, two different methods of calculation must lead to the same 
results. However, the values of ${\rm Re}{\zeta}_{22}^{(ef)}/{\rm 
Re}\zeta$ for deep ($h/2a>1$) grooves obtained in the present work (the 
stars in Fig.3a) are approximately twenty percents less than the 
results of \cite{17}. Trying to find out the origin of this discrepancy 
and to verify our approach, in \cite{18} we examined the system 
investigated in \cite{17} using two different approaches. We reproduced 
the perturbation theory calculations of \cite{17} with the aid of the 
modal functions method used in the present work. The results were the 
same as in \cite{17}. Then we repeated the calculation with the aid of 
Eq.(27.3). Comparing the results we obtained the same twenty percents 
difference. The analysis of \cite{18} showed that the only source of the 
discrepancy could be in application of the perturbation theory. 

Unfortunately, we were not successful in detecting accurately why the 
standard perturbation theory used in \cite{17} and \cite{18} led to the 
results differing from the results obtained with the aid of our approach 
based on Eq.(27.3). However, we found some reasons allowing us to assume 
that in this problem the standard perturbation theory was not justified 
properly (see \cite{18}). We would like to note that the aforementioned 
difference is not of principle for qualitative description of the 
results obtained when analyzing the reflection of s-polarized waves from 
lamellar gratings.

\section{SUMMARY}

On the basis of the local impedance boundary conditions (1) we examined 
several metallic systems with strong surface inhomogeneities. For each 
system we calculated the effective surface impedance tensor. Our results 
are exact within the accuracy of Eqs.(1).

If the surface of an inhomogeneous conductor is flat, the effective 
impedance is equal to the values of the local impedances averaged over 
the surface (see Eq.(7)). As an example, we examined a flat metallic 
surface with one-dimensional periodic strip-like local surface 
impedance. In this case the effective impedance is an isotropic tensor: 
${\zeta}_{ik}^{(ef)}= {\zeta}_{ef}{\delta}_{ik}\,(i,k = 1,2)$; 
${\zeta}_{ef}$ is defined in Eq.(10). 

We analyzed the role of evanescent waves paying special attention to the 
difference between the p- and s-polarization states. We showed that in 
the p-polarization state the evanescent waves did not contribute to the 
magnetic field in the immediate vicinity of the surface. On the 
contrary, in the s-polarization state just the evanescent waves provide 
the main contribution to the magnetic field taking account of the finite 
conductivity. 

For this problem we also calculated the components of the time-averaged 
Pointing vector $\overline{\bf S}$. For normally incident waves the 
force lines of the vector $\overline{\bf S}$ being directed along the 
$x_3$ axis far from the surface ($x_3 > \lambda$) become distorted near 
the surface. The stronger is the surface inhomogeneity, the more the 
distortion of the lines. We showed that in the p-polarization state the 
Pointing vector was defined by the complex potential $w(z)$ ($z = x_1 + 
ix_3$) (see Eq.(19)). The aforementioned difference in the contribution 
of evanescent waves to the magnetic field is the reason why a complex 
potential cannot be introduced in the s-polarization state. 

When the inhomogeneity is very strong (narrow strips with rather large 
impedance separated by wide strips with very small impedance), for both 
polarizations all the force lines of the vector $\overline{\bf S}$ meet 
at the surface in the narrow regions where the local impedance is big 
(see Fig.1). In the p-polarization state the equation for these lines is 
just the same as the equation for the force lines of the vector $\bf E$ 
in the electrostatic problem for the periodic system of charged straight 
lines. 

As an example of calculation of the effective impedance in the case of a 
conductor with rough surface we analyzed a periodic 1D surface of 
homogeneous isotropic conductor depicted in Fig.2a. In this case the 
effective impedance is a tensor whose elements differ significantly, 
when the grooves are rather deep. 

In the p-polarization state the electromagnetic field penetrates into 
the grooves on the surface. In our frequency region this means that the 
magnetic field $H_2^{per}$ is nearly constant in the immediate vicinity 
of the surface. As a result, the surface as a whole takes part in 
absorption of p-polarized waves. This leads to a significant enlargement 
of absorption: the longer the length of the surface profile, the more 
the real part of the element of the effective impedance tensor relating 
to the p-polarization state (see Eq.(31)). In the framework of our 
approximation the upper bound for the real part of this element of the 
effective impedance tensor is of the order of one. The surfaces showing 
the giant absorption of p-polarized waves (${\zeta}_{ef} \sim 1$) can be 
realized, for example, if the contour of the grooves is a strongly 
branching line.

In the s-polarization state evanescent waves contribute to the magnetic 
field in the vicinity of the surface. As a result at the surface the 
components of the magnetic vector ${\bf H}^{per}$ depend on coordinates. 
The most important that inside the grooves all the fields decay 
exponentially with the increasing distance from the plane $x_3=0$. No 
simple expressions for this element of the effective impedance tensor 
can be obtained. For each surface the calculation has to be done 
separately beginning from the calculation of magnetic field near the 
infinitely conducting surface of the same geometry.

As an example we examined the lamellar gratings shown in Fig.2c. 
Recently many authors studied scattering from surfaces having 
rectangular grooves. These surfaces are of interest because they can be 
manufactured very easily providing a possibility to check theoretical 
results. Most of the investigations have been devoted to resonant 
enhancement processes (see, for example, \cite{12,13,16}). However, the 
frequency region examined in these works does not comply with the 
inequalities (2) defining the framework of the effective impedance 
approach applicability.  

Our numerical results obtained when calculating the effective impedance 
associated with the lamellar grating for the s-polarization state are 
shown in Figs.3 and Fig.4. 

Although in the s-polarization state the presence of the grating leads 
to an increase of the real part of the effective impedance too, even for 
very deep grooves it cannot achieve such giant values that are possible 
in the p-polarization state. Taking into account that in the case of an 
arbitrary 1-D surface profile s-polarized waves do not penetrate into 
deep grooves, we can generalize our result: if an arbitrary 1-D surface 
profile has rather deep grooves, the effective impedance associated with 
this surface is a strongly anisotropic tensor. Its element relating to 
the p-polarized state is much greater than the element relating to the 
s-polarized state.
\vskip 3mm

{\bf ACKNOWLEDGEMENTS}

The authors are gratefully to prof. M.I.Kaganov and dr. T.A.Leskova for 
helpful discussions. The work of IMK was supported by RBRF grant 99-02-
16533.

\section{APPENDIX 1}

Herein for the p-polarization state we calculate the leading term in the 
expression for the total magnetic field ${H}_{2}^{per}$ above the 
infinitely conducting lamellar grating depicted in Fig.2c supposing that 
$h,b \ll \lambda$.

In the half-space above the grooves $ (x_3>0)$ we adopt the plane-wave 
representation of the magnetic field ${H}_{2}^{per} = 
{H}_{2}^{+}({x}_{1},{x}_{3})$: 
$$
{H}_{2}^{+}({x}_{1},{x}_{3}) = {H}_{0}\left\{
{\rm e}^{-ikx_3} + \sum_{q = -\infty}^{\infty}
{h}_{q}^{+}{\rm e}^{i(\pi q{x}_{1}/b + {\alpha}_{q}{x}_{3})}\right\},
\eqno (A.1)
$$
${\alpha}_{q}$ is defined in Eq.(34.2); $H_0$ is the amplitude of the 
incident wave.

Inside the central groove ($|x_1|<a; -h<x_3<0$) we seek ${H}_{2}^{per} = 
{H}_{2}^{-}({x}_{1},{x}_{3})$ as a series of modal functions 
${\phi}_{n}^{(p)}(x_1,x_3)$ that are the solutions of the Maxwell 
equations. With regard to the boundary conditions for perfect conductors 
we have 
$$
{H}_{2}^{-}({x}_{1},{x}_{3}) ={H}_{0}\sum_{n=0}^{\infty}{B}_{n}^{(p)}
{\phi}_{n}^{(p)}(x_1,x_3); \quad {\phi}_{n}^{(p)}(x_1,x_3) =
\cos \left(\frac{\pi n}{2a}({x}_{1} - a)\right)\cos {\beta}_{n}(h + {x}_{3}),
\eqno (A.2)
$$
${\beta}_n$ is defined in Eq.(34.1).

Applying the boundary conditions on the plateaus $x_3=0$ and the 
continuity conditions across the central slit, we obtain the set of 
equations for the coefficients ${B}_{n}^{(p)}$: 
$$
i\sum_{n=0}^{\infty}({\beta}_{n}b){B}_{n}^{(p)} \sin 
({\beta}_{n}h){I}_{nm}^{(p)}
- \frac{a}{2b}[{B}_{m}^{(p)}\cos {\beta}_{m}h + 
{\delta}_{m0}{B}_{0}^{(p)}\cos (kh)] = 
-2{C}_{m0}^{*},
\eqno (A.3a)
$$
$$
{I}_{nm}^{(p)}=\sum_{q=-
\infty}^{\infty}\frac{1}{({\alpha}_{q}b)}{C}_{nq}{C}_{mq}^{*},
\eqno (A.3b)
$$
$$
{C}_{nq} = 2i\frac{a}{b}A(m; 2qa/b), \; A(m;z) = \frac{z}{2\pi} 
\frac{[{\rm e}^{-i\pi z/2} - {(-1)}^{m}{\rm e}^{i\pi z/2}]} 
{[{z}^{2}-m^2]},
\eqno (A.3c)
$$

In terms of the coefficients ${B}_{n}^{(p)}$ the amplitudes 
${h}_{q}^{+}$ are:
$$
{h}_{q}^{+} = {\delta}_{q0} + \frac{i}{{\alpha}_{q}}\sum_{n=0}^{\infty}
{B}_{n}^{(p)}{\beta}_{n}{C}_{nq}\sin {\beta}_{n}h.
\eqno (A.4)
$$
When the elements of the matrix ${I}_{nm}^{(p)}$ are presented as the 
series expansions in powers of the small parameter $\gamma = kb$, it is 
easy to see that the element ${I}_{00}^{(p)}$ has the term $a/2b\gamma$, 
and the expansions of all the other elements begin with the terms 
independent of $\gamma$. Then from Eqs.(A.3a) it follows that up to the 
terms independent of $\gamma$ 
$$
{B}_{0}^{(p)} = 2; \quad {B}_{n}^{(p)} = 0, \;{\rm if}\, n \ne 0.  \eqno 
(A.5a)
$$ 
Within the same accuracy from Eq.(A.5) we have
$$
{h}_{0}^{+} = 1; \quad {h}_{q}^{(+)} = 0, \;{\rm if}\, q \ne 0.  \eqno 
(A.5b)
$$
Thus, inside the grooves the modal function with $n=0$ gives the main 
contribution to the magnetic field. Next, the amplitudes of evanescent 
waves are negligibly small in comparison with the amplitude of the 
reflected wave. With regard to Eq.(A.1) and Eq.(A.2) this means that in 
the vicinity of the surface the magnetic field is nearly constant: 
${H}_{2}^{per}({x}_{1},{x}_{3}) \approx 2H_0$.

\section{APPENDIX 2}

In the s-polarization state when calculating the components of the 
magnetic vector ${\bf H}^{per}$, special attention has to be drawn to 
the corner points of our lamellar grating. It is well known that if a 
surface has geometrical singularities, some components of 
electromagnetic vectors have singularities two. In the s-polarization 
state the tangential magnetic field in the immediate vicinity of a 
rectangular two-dimensional infinitely conducting wedge (the wedge is 
along the $x_2$ direction) is proportional to ${\rho}^{-1/3}$, where 
$\rho = \sqrt{x_1^2+x_3^2}$ is the distance from the corner point (see, 
for example, \cite{15}). Consequently, ${H}_{1}^{(+)}(x_1,0) \sim {|x_1 
\pm a|}^{-1/3}$, when $x_1 \to \pm a$, and ${H}_{3}^{(-)}(\pm a, x_3)$, 
when $x_3 \to 0$. We need the solution of Eqs.(36) providing the 
aforementioned behavior of the magnetic field.

In \cite{15} it was shown that the behavior of the fields near the edge 
is defined by the asymptotic behavior of the modal functions 
representing the fields. Let us show that in our case the tangential 
magnetic field near the edge is proportional to ${\rho}^{-1/3}$,  if 
$Y_n \sim 1/{(2n-1)}^{2/3}$, when $n \to \infty$.

Indeed, let us examine, for example, ${H}_{3}^{(-)}(x_1,x_3)$ when 
$x_1=a$ and $x_3 \to -0$. It is the tangential magnetic field on the 
vertical facet of the groove near its throat. Taking into account that 
when $x_3 \to 0$, the behavior of ${H}_{3}^{(-)}(a,x_3)$ is defined by 
the coefficients $Y_n$ with $n \to \infty$, with regard to Eq.(37.4) we 
obtain
$$
{H}_{3}^{(-)}(a,x_3) \approx \frac{\pi 
b}{2a}\sum\limits_{n=1}^{\infty}Y_n {\rm e}^{-n|x_3|}, \quad |x_3| \to 
0.   \eqno (A.6a)
$$  
Suppose $Y_n \sim {(2n-1)}^{p}$ when $n \to \infty$. With regard to the 
asymptotic equality \cite{15}
$$
\sum\limits_{n=1}^{\infty}n^p {\rm e}^{-nz} \approx \Gamma (1+p){z}^{-
(p+1)}, \quad z \to +0 \; {\rm and}\; -1<p<0,   \eqno (A.6b)
$$
($\Gamma (x)$ is the Gamma function), comparing equations (A.6a) and 
(A.6b) we find that $p = -2/3$ has to be chosen. 

The next step is to search the solution of the set (36.1) truncating it 
to finite order N and increasing N until stability of successive 
solutions is obtained. Simultaneously, the sum defining the matrix 
${\Delta}_{nm}$ (see Eq.(36.2) has to be truncated to the finite order 
$Q$. The solution has to guarantee that $Y_n\sim {(2n-1)}^{-2/3}$, when 
$n \gg 1$.

In \cite{15} a similar set of equations was analyzed. This set was 
obtained when calculating the fields inside a rectangular one-
dimensional branching wave-guide. It was shown that such set of 
equations allowed unrestricted number of solutions depending on the 
limit of the ratio $Q/N$ for $Q,N \to \infty$. The value of this limit 
ensuring the necessary behavior of the fields near the edge of the wedge 
determined the physically meaningful solution of the set in question. 
For a branching wave-guide this limit was calculated analytically as a 
function of the ratio of the sizes characterizing the wave-guide.
 
Unfortunately, in the case of the lamellar grating we failed trying to 
define analytically the limit $\lim_{N,Q \to \infty}(Q/N)$ providing 
$Y_n \sim {(2n-1)}^{-2/3}$ for $n \to \infty$. Our results are based on 
numerical calculations. 

In all the calculations we used $N=40$. For this truncation number the 
stabilization was obtained in all examined examples. This value of $N$ 
was sufficient to obtain a stable exponential solution 
$Y_n \sim {((2n-1)}^{-\tau}$ for $n \gg 1$ as well as the stable results for 
the real and imaginary parts of the effective impedance calculated with the 
aid of (33.1) and (38).

To determine the ratio $Q/N$ for given values of $h/a$ and $a/b$, we 
solved the set of equations (36) using different truncation numbers $Q$ 
and chose $Q$ ensuring $\tau = 2/3$. We find that the ratio $Q/N$ 
providing the given value of $\tau$, depends on the ratio $h/a$ only 
slightly. When calculating $Q/N$ as function of $a/b$ we saw that rather 
accurately the value $\tau =2/3$ corresponded to the constant value of 
the ratio $aQ/bN$. For $h/a = .2$ and $h/a = 2$ the values of $aQ/bN$ 
were 1.9 and 2.1 respectively. In Fig.5 for $a/b = 1/sqrt(2)$ and $h 
= 2a$ we show the exponent $\tau$ as function of $x$: $Q = [xbN/a]$ 
($[z]$ denotes the integral part of the number $z$).

It worth to be mentioned that although for $n \gg 1$ the behavior of the 
coefficients $Y_n$ depends on the $\lim_{N,Q \to \infty}(Q/N)$, for 
small numbers $n$ the coefficients $Y_n$ are practically the same when 
$Qa/Nb >1$.

\newpage
\centerline{\bf LIST OF FIGURES}

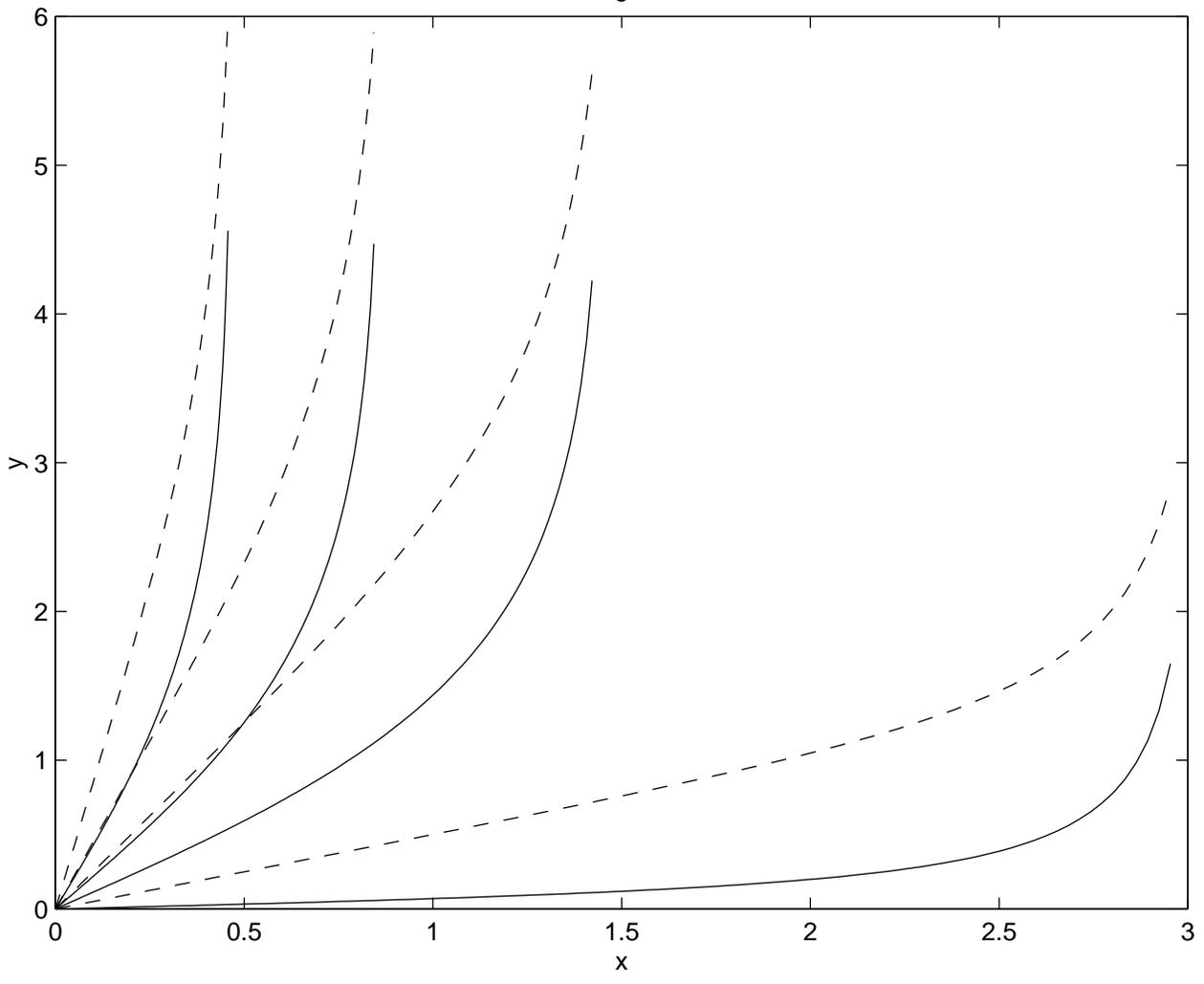
Fig.1. The force lines of the time-averaged Pointing vectors 
$\overline{{\bf S}^{(p)}}$ (full lines) and $\overline{{\bf S}^{(s)}}$ 
(dashed lines) near the flat metallic surface with strongly anisotropic 
strip-like local impedance.

Fig.2a. One-dimensional periodic surface that is examined in Section 4.

Fig.2b. A branching contour providing giant absorption of the incident 
p-polarized wave.  

Fig.2c. The lamellar grating configuration used to calculate the 
effective impedance in the s-polarization state.

Fig.3a. The ratio ${\rm Re}{\zeta}_{22}^{(ef)}/ {\rm Re}\zeta$ versus 
$a/b$ for $h = 10a$ (the stars) and $h = 0.2a$ (the dots). 

Fig.3b. The values of $Z$ versus $a/b$ for $h = 10a$ (the stars) and $h 
= .2a$ (the dots).

Fig.3.c. The ratio $Z_{pl}/Z_s$ versus $a/b$ for $h = 10a$ (the stars) 
and $h = .2a$ (the dots).

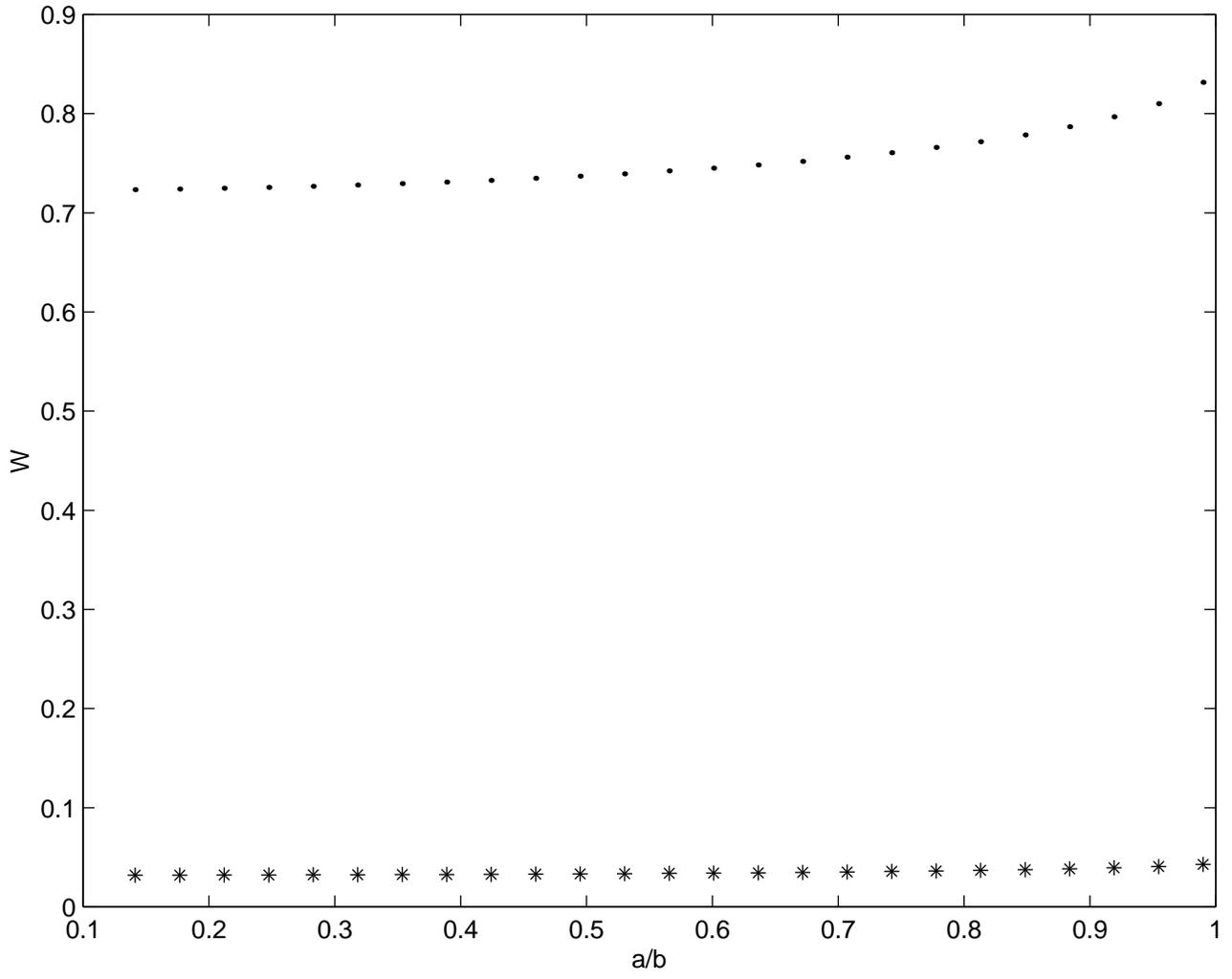
Fig.4. The function $W$ versus $a/b$ for $h = 10a$ (the stars) and $h = 
.2a$ (the dots). 

Fig.5 The exponent $\tau$ as function of $Q/N$ of $x$ for $h=2a$ and 
$a/b=1/sqrt(2)$ ($Q = [xbN/a]$).

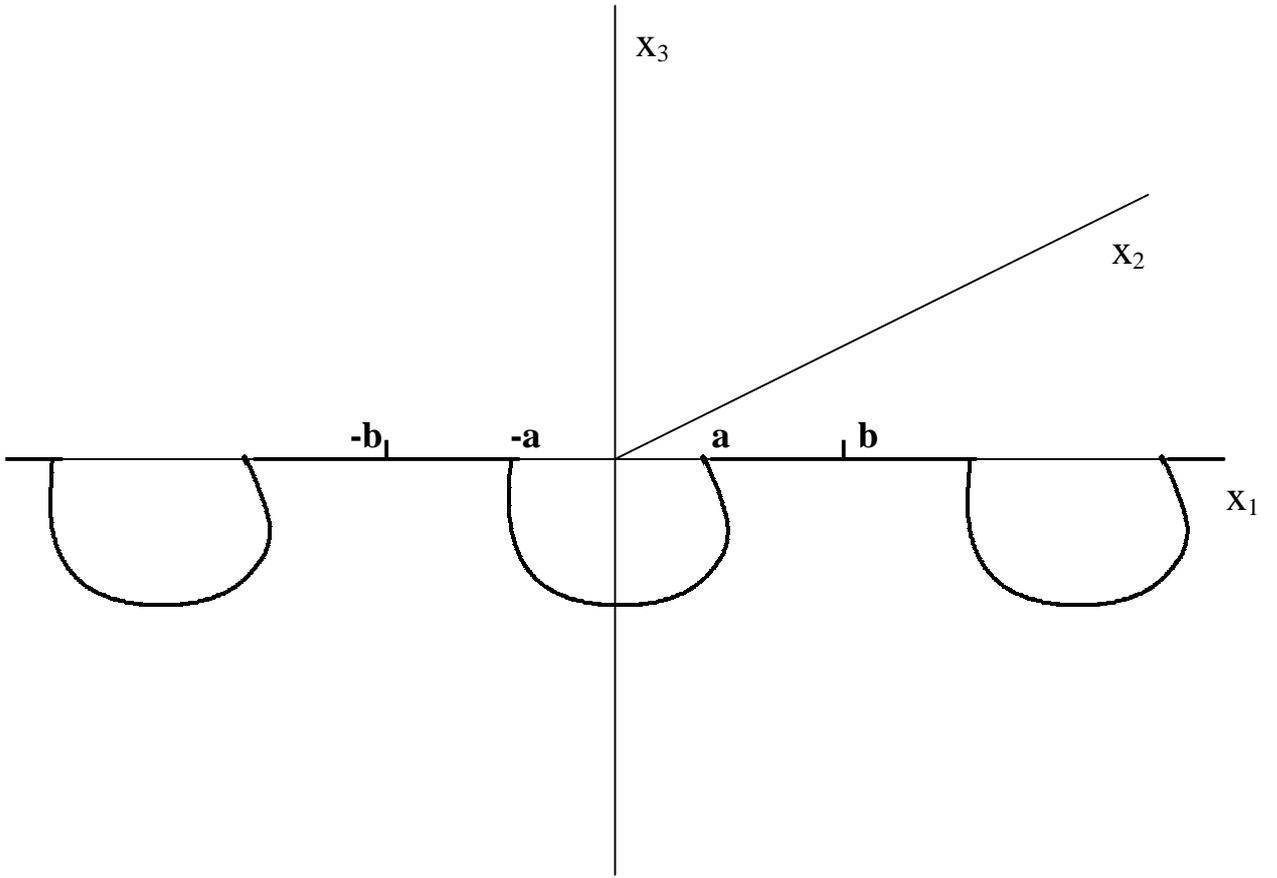
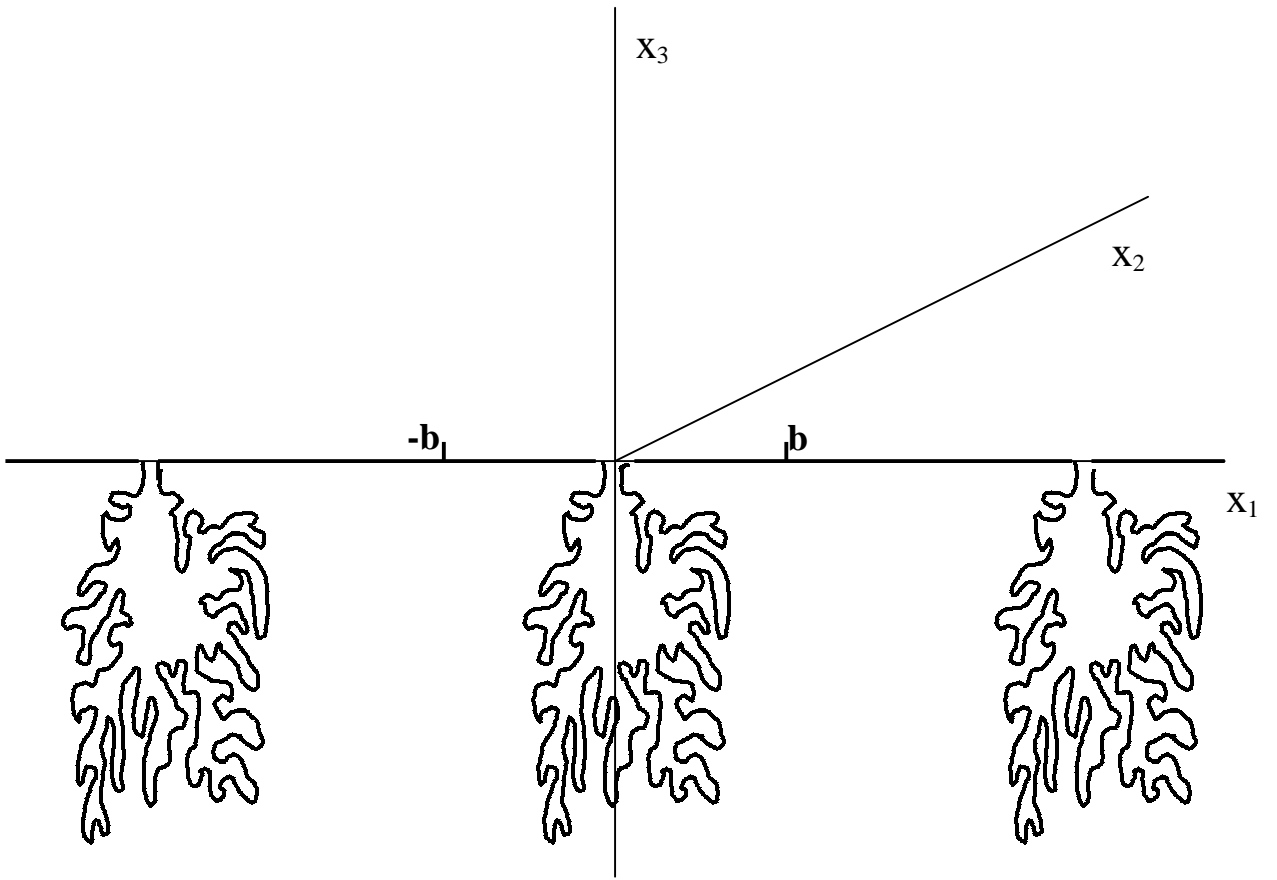
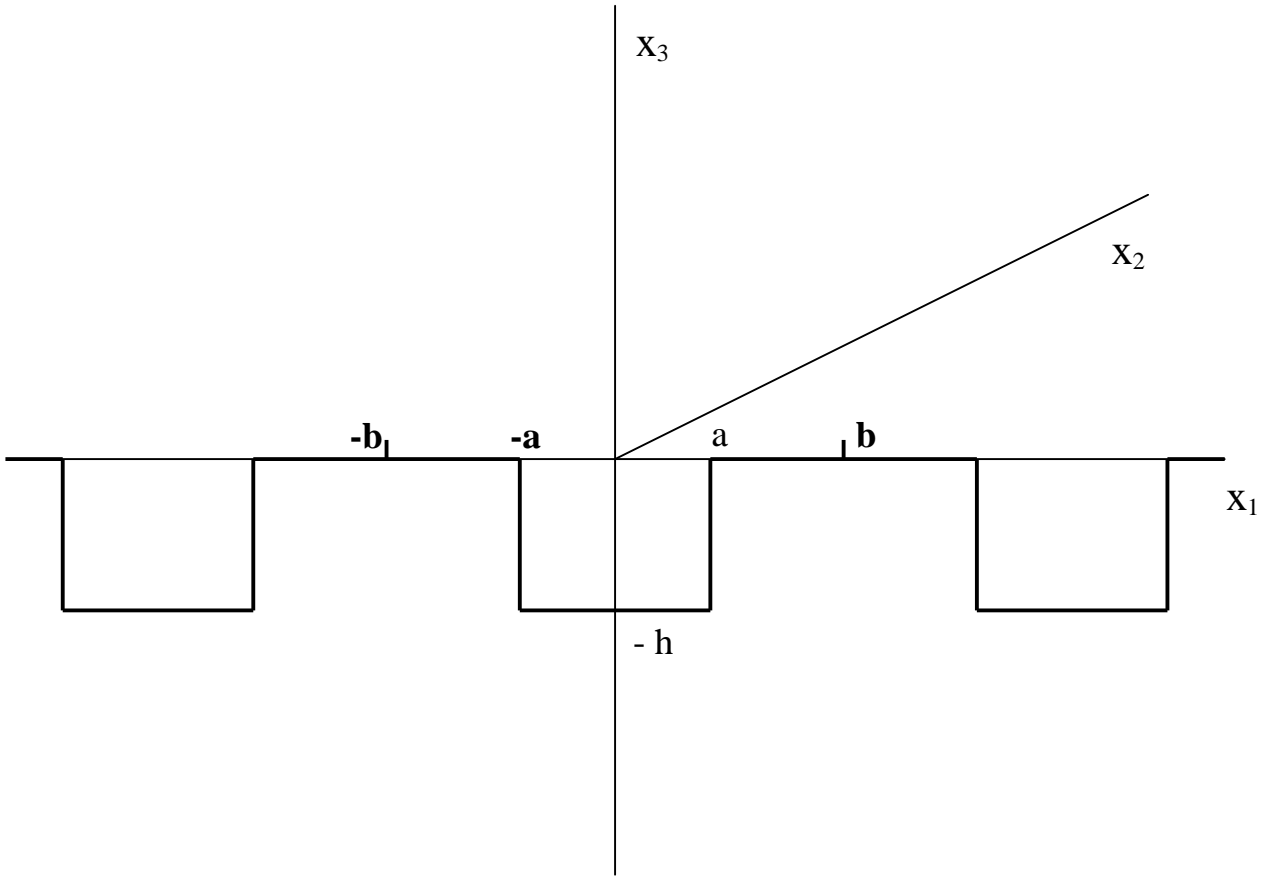
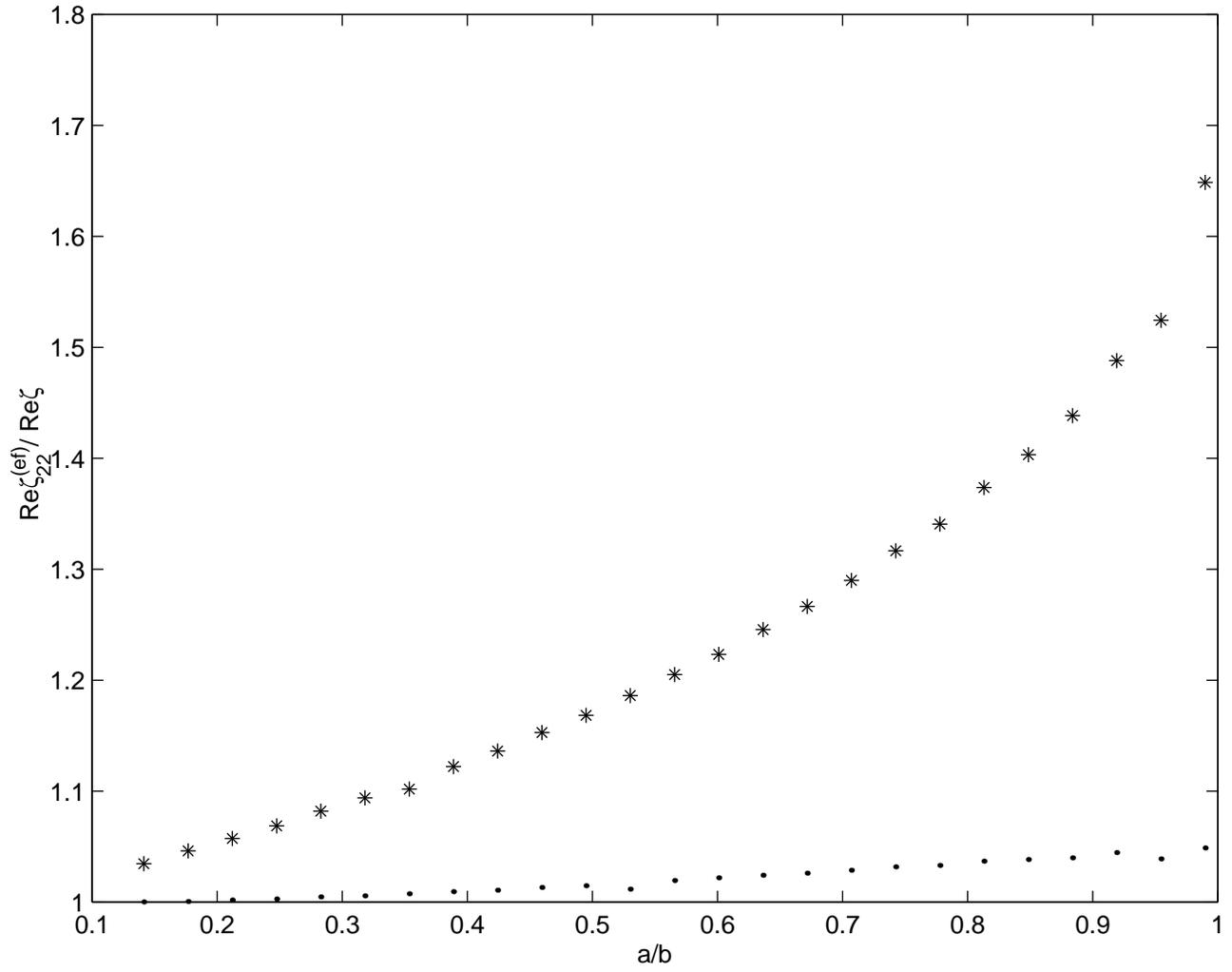
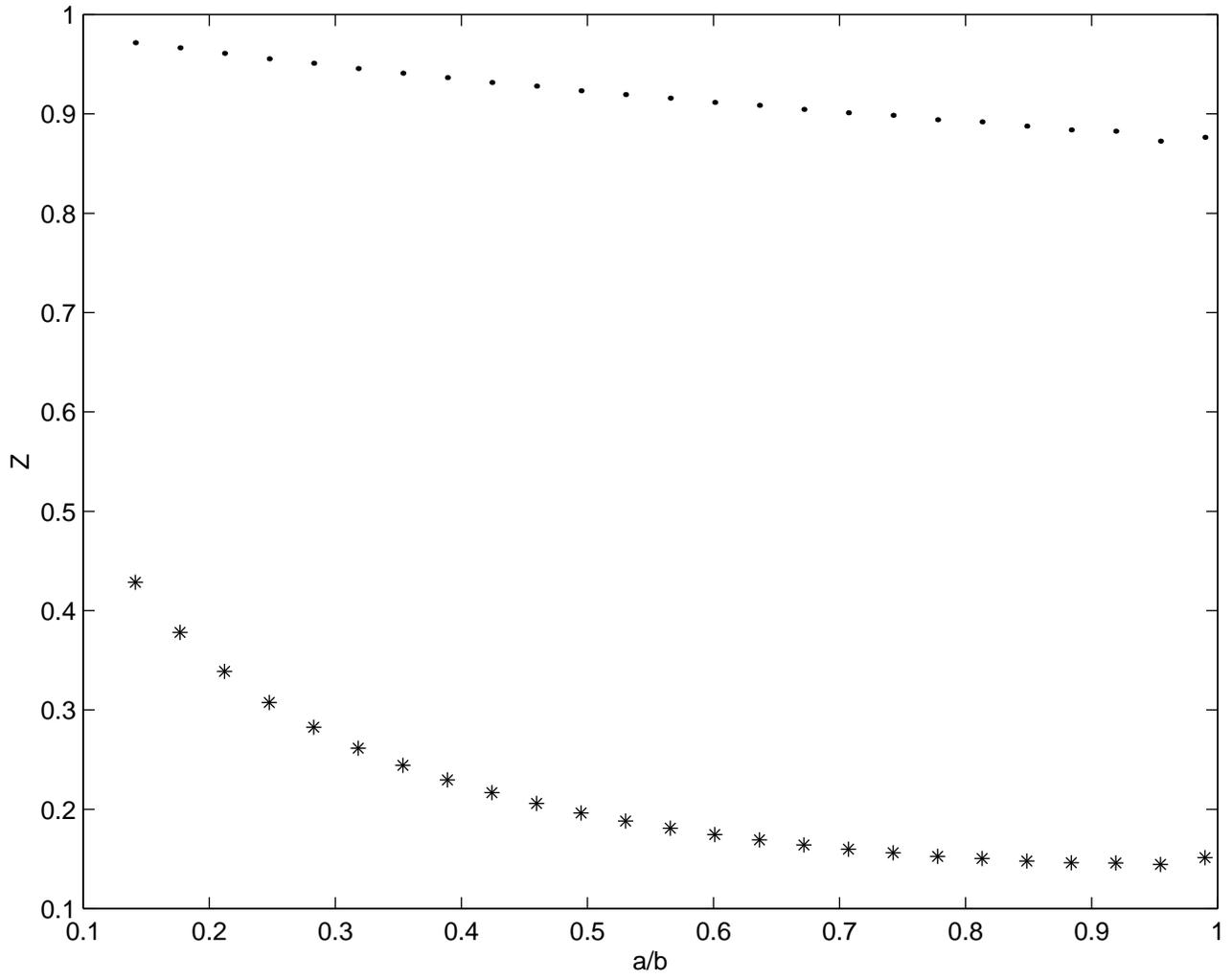
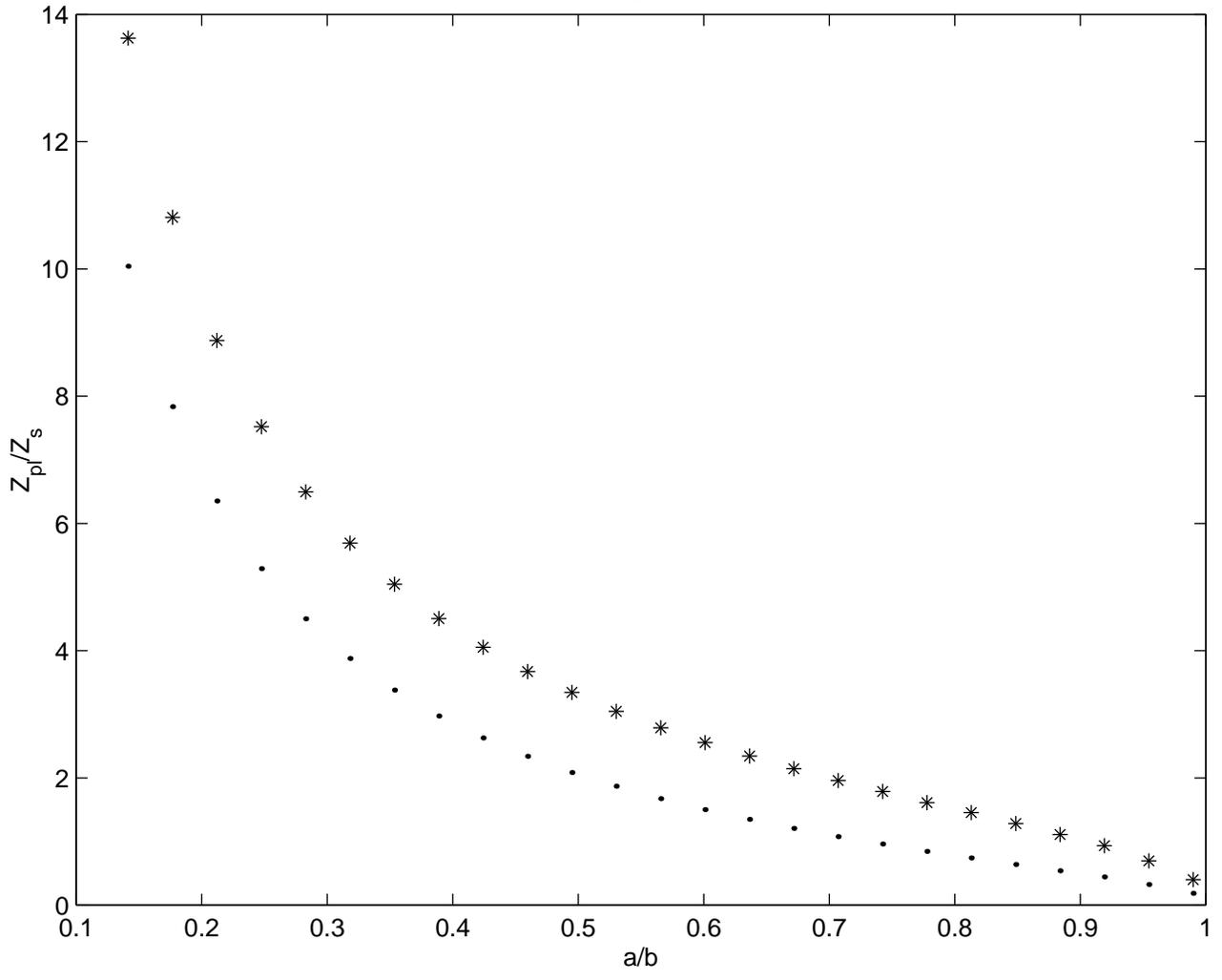
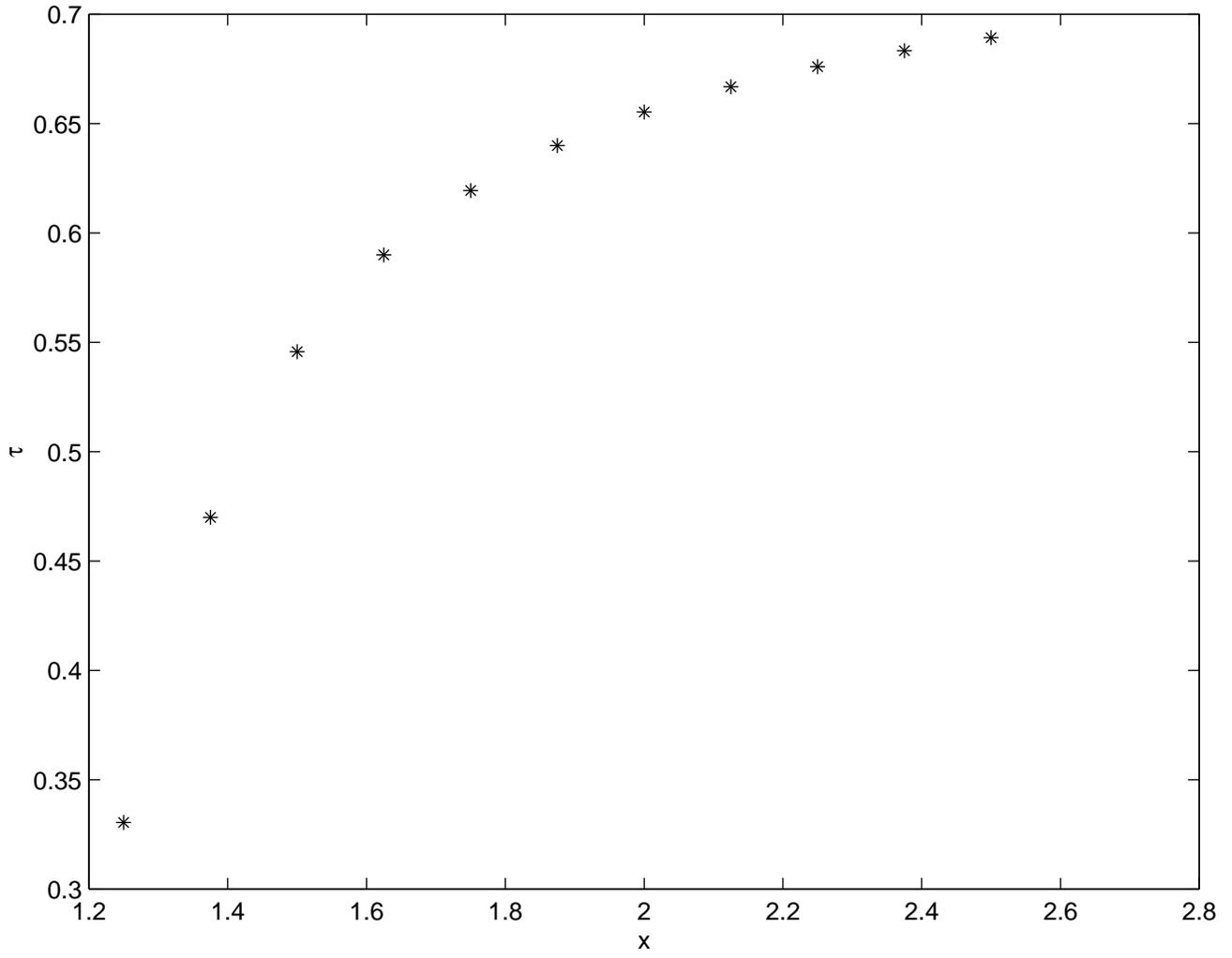
\end{document}